\documentclass[twocolumn,showpacs,preprintnumbers,amsmath,amssymb,a4paper,pre]{revtex4}
\usepackage{graphicx}
\usepackage{amssymb, amsmath,amssymb,amsfonts}
\usepackage{graphicx}
\usepackage{dcolumn}
\usepackage{bm}
\usepackage{subfigure}
\usepackage{color}
\begin{document}
\title{Diffusion of tagged particles in a crowded medium}
\author{Marta Galanti$^{1,2,4}$, Duccio Fanelli$^{2}$, Amos Maritan$^{3}$, Francesco Piazza$^{4}$}
\affiliation
{\mbox{$^{1}$ Dipartimento di Sistemi e Informatica,Universit\`a di Firenze and INFN}, \\
\mbox{Via S. Marta 3, IT-50139 Florence, Italy}\\
\mbox{$^{2}$ Dipartimento di Fisica e Astronomia, Universit\`a di Firenze and INFN}, \\
\mbox{Via Sansone 1, IT-50019 Sesto Fiorentino, Italy}\\
\mbox{$^{3}$ Dipartimento di Fisica e Astronomia "G. Galilei" Universit\`a di Padova, CNISM and INFN}, \\
\mbox{via Marzolo 8, 35131, Padova, Italy}\\
\mbox{$^{4}$ Universit\'{e} d'Orl\'eans and Centre de Biophysique Mol\'eculaire (CBM),} \\
\mbox{CNRS-UPR 4301, Rue Charles Sadron, 45071 Orl\'eans, France}
}
\begin{abstract}
The influence of crowding on the diffusion of tagged particles in a dense medium is investigated in the framework of a 
mean-field model, derived in the continuum limit from a microscopic stochastic process with exclusion. 
The probability distribution function of the tagged particles obeys to a nonlinear Fokker-Planck equation, 
where the drift and diffusion terms are determined self-consistently by the concentration of crowders in the medium. 
Transient sub-diffusive or super-diffusive behaviours are observed, depending on the selected initial conditions, 
that bridge normal diffusion regimes characterized by different diffusion coefficients. 
These anomalous crossovers originate from the microscopic competition for space and reflect 
the peculiar form of the non-homogeneous advection term in the governing Fokker-Planck equation. 
Our results strongly warn against the overly simplistic identification of crowding with anomalous transport {\em tout court}. 
\end{abstract}
%
%%%%%%%%%%%%%%%%%%%%%%%%%%%%%%%%%%%%%%%%%%%%%%%%%%%%%%%%%%%%%%%%%%%%%%%%%%%%%%%%%%%%%%%%%%%%%%%%%%%%%%%%%%%%%%%%%%%%%%%%%%%%%%%%%
\maketitle
%
%%%%%%%%%%%%%%%%%%%%%%%%%%%%%%%%%%%%%%%%%%%%%%%%%%%%%%%%%%%%%%%%%%%%%%%%%%%%%%%%%%%%%%%%%%%%%%%%%%%%%%%%%%%%%%%%%%%%%%%%%%%%%%%%%
%
\section{Introduction}
%
%%%%%%%%%%%%%%%%%%%%%%%%%%%%%%%%%%%%%%%%%%%%%%%%%%%%%%%%%%%%%%%%%%%%%%%%%%%%%%%%%%%%%%%%%%%%%%%%%%%%%%%%%%%%%%%%%%%%%%%%%%%%%%%%%

\noindent Diffusion is a fundamental process in nature that describes the spread of particles subject to random forces 
from regions of high density to regions of low density~\cite{Crank:uq}.  The hallmark of diffusive transport 
is the linear growth in time of the 
mean square displacement (MSD) of the spreading particles, $\langle \Delta R^{2}\rangle \propto t$. 
This is a simple conclusion that follows directly from the law of conservation 
of matter (in the form of a continuity equation), when a simple constitutive equation is assumed,
stating that the particle current is proportional to the concentration gradient. The latter law,
known as the (first) Fick's law, can be regarded as a simple linear-response prescription, thus 
only appropriate to describe the relaxation of small density fluctuations.\\
\indent Despite the fact that Fickean diffusion is generally appropriate to describe the spontaneous spatial
rearrangement of particles in suspension, deviations are expected to occur in various situations of interest,
{\em e.g.} if fixed obstacles are present (confinement) \cite{Bressloff}  or when different, and thus distinguishable
species compete for the available space 
at high concentration, a scenario often referred to in cellular biology as 
{\em macromolecular crowding}~\cite{Foffi:2013kr,Rivas:2004em,Dix:2008lp,Ellis:2001bu,Ellis:2001wv,Konopka01092006,Minton:2005dm,
Schnell:2004dd,Weiss:2004vn,Zhou:2004km,Zhou:2008vf,Szymanski:2006jg,Phair:2000wa,Piazza:2012ik}. \\
\indent Despite the importance of crowding  and confinement effects in diffusion-related mechanisms in chemistry and 
biology, there is no consensus on the mechanisms through which crowding and confinement fine-tune deviations 
from the classical Fickean picture.
This lively debate is reflected by conflicting experimental reports in the literature concerning 
the role of complex environmental factors in the mobility of biomolecules in the cytoplasm and extra-cellular matrix. 
Some authors maintain that crowding  merely slows down transport by reducing in a complex fashion 
the diffusion coefficient but does not alter the MSD exponent~\cite{Dix:2008lp,Novak:2009uq,Dauty:2004hb,Szymanski:2006jg},
while others~\cite{Golding:2006ew,Pastor:2010ee,Weiss:2004vn,Banks:2005cc} contend the identification of crowding in the cytoplasm
with anomalous (typically sub-diffusive) transport~\cite{Bouchaud:1990ys,Metzler:2000du,Zaslavsky:2002wn}, a feature  
observed in lateral diffusion in cellular membranes~\cite{Javanainen:2012km,Kneller:2011wf,Feder:1996gm}.
In this case one would have $\langle \Delta R^{2}\rangle \propto t^{\alpha}$ with $\alpha<1$ (sub-diffusion) or
$\langle \Delta R^{2}\rangle \propto t^{\alpha}$ with $\alpha>1$ (super-diffusion). 
It is worthwhile to underline that reports of anomalous transport connected to crowding are not limited to sub-diffusion.
For example, Upadhaya and collaborators~\cite{Upadhyaya:2001gk} have recorded super-diffusive behaviour in 
the motion of endodermal Hydra cells, which they traced back to long-range correlations 
within the scrutinized sample, while Stauffer and collaborators~\cite{Stauffer:2008bx} 
proposed a minimalistic model of random barriers in a percolation network as a tool 
to mimic diffusion in a crowded environment.\\
\indent To add an important piece of information to the debate, it is interesting to remark that most often 
claims of anomalous diffusion in three-dimensional crowded environments {\em in vitro}
and {\em in vivo} rely on fluorescence recovery after photobleaching (FRAP) data that 
are analyzed through {\em ad hoc} modifications~\cite{Pastor:2010ee} of standard theories of 
fluorescence photobleaching recovery~\cite{Soumpasis:1983kr,Axelrod:1976iy}.
It is interesting to remark that to our knowledge 
no first-principle derivations of fluorescence recovery curves in the anomalous diffusion regime 
have yet been reported, analogous to the long-known standard derivations performed 
in the context of normal diffusion~\cite{Soumpasis:1983kr,Axelrod:1976iy}.\\
\indent As it is often the case, the truth probably reflects an intermediate picture. Possibly, 
complex (even multiple) crossovers are to be expected between anomalous and normal diffusion~\cite{Vilaseca:2011fq},
or, alternatively, one needs to consider complex space- and geometry-dependent diffusion 
coefficients~\cite{Zador:2008ie,Nicholson01121981}, as modeled {\em e.g.} by Fick-Jacobs~\cite{Jacobs:1967fk} and 
related theories~\cite{Martens:2013tp,Kalinay:2013il}. 
However, as it appears clear from the above recollection, the need for further, systematic investigation 
of transport in crowded and confining media is evident.\\
\indent A particularly interesting approach to model transport in complex media is to derive  macroscopic  
equations as mean-field approximations of suitable microscopic stochastic processes. In this way, 
the microscopic constraints imposed by complex environmental factors are {\em naturally} incorporated 
in the transport equations~\cite{Nossan:2013fk,Schonherr:2004zr,Schutz:1993ye,Basu:2010pi,Simpson:2009ys}. 
For example, in Ref.~\cite{Galanti:2013gb} we derived a modified nonlinear  equation suitable
for describing the mean-field limit of a persistent random walk in a dense environment. \\
\indent The idea is to move from a  space-discrete simple 
exclusion process specifying the competition for space at the microscopic level. 
This is an agent-based stochastic model bound to the condition that no two agents can  occupy 
the same site~\cite{Privman:2005wo,Liggett:1999vp,Derrida:1998oq}. In certain limits, the governing equations obtained through such procedure 
can be also viewed as nonlinear Fokker-Planck equations derived from a master equation or from generalized 
free-energy functionals~\cite{Chavanis:2008fb}. \\
\indent In this paper we consider the diffusion of tagged particles immersed in a densely populated {\em milieu}
of co-evolving agents, hereafter the {\em crowders}, as a primer for most fluorescence-based single-molecule tracking
experiments.
Following an approach inspired from Ref.~\cite{Galanti:2013gb}, 
we derive a system of partial differential equations for the mean-field densities of both the tagged particles and the 
crowders. The model is formulated at the microscopic level as a stochastic process with simple exclusion interferences. 
In the thermodynamic limit, excluded-volume constraints result in nonlinear coupling terms between the two concentrations.
We observe that this is expected, as multicomponent diffusion should be in general nonlinear 
if there exist non-diagonal terms, for diffusion preserves the positivity of particle densities~\cite{Gorban:2011fw}.\\
%
%%%%%%%%%%%%%%%%%%%%%%%%%%%%%%%%%%%%%%%%%%%%%%%%%%%%%%%%%%%%%%%%%%%%%%%%%%%%%%%%%%%%%%%%%%%%%%%%%%%%%%%%%%%%%%%%%
%
%  FIGURA 1 
%
\begin{figure}[t!]
\centering
\includegraphics[width=\columnwidth,clip]{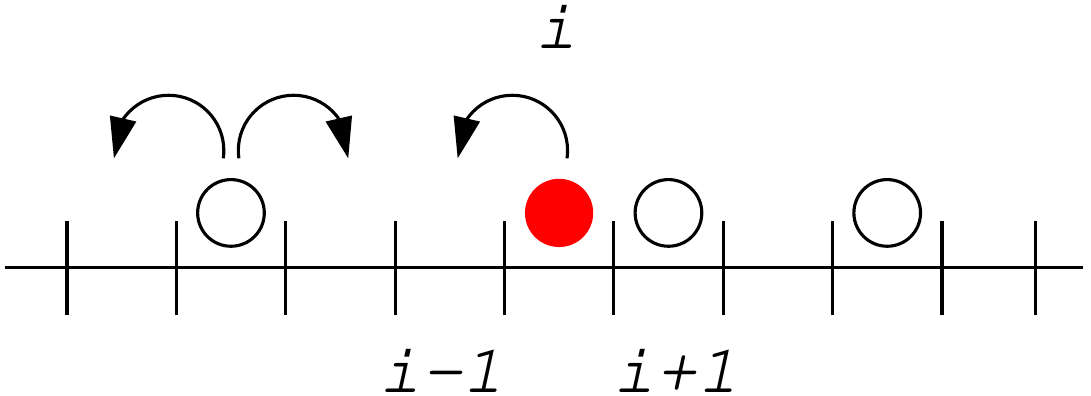}
\caption{The simple exclusion rule implemented in the model of tagged particle diffusion. 
In this configuration, the tagged particle (filled circle) sitting at site  $i$ can only jump 
towards the (empty) neighbor site $i-1$. The jump towards site $i+1$ is instead impeded, as the 
target site is occupied by a crowder (empty circles). Crowders can also diffuse towards neighbors sites. 
\label{fig:exclud}}
\end{figure}
%%%%%%%%%%%%%%%%%%%%%%%%%%%%%%%%%%%%%%%%%%%%%%%%%%%%%%%%%%%%%%%%%%%%%%%%%%%%%%%%%%%%%%%%%%%%%%%%%%%%%%%%%%%%%%%%%
%
\indent Our paper is organized as follows. In section~\ref{sect:1} we introduce our model and work out the system of coupled
mean-field transport equations. In section~\ref{sect:2}, 
we study the spreading of an initially localized collection of tagged particles with different starting configurations
of the crowders as possible realizations of feasible and interesting experiments. Remarkably, we show that transient 
sub-diffusion and super-diffusion can be observed, depending on the specific initial conditions, as crossovers
between two normal diffusion regimes with different diffusion constant. We rationalize our findings by showing how
the initial condition impacts on the sign of an effective advection 
term in the Fokker-Planck equation for the tagged particles. In section~\ref{sect:3} we summarize our results and 
stress the important conclusions reported in this paper.

%=================================================================================================================
%  SECTION 1
%
\section{The microscopic model and its mean-field limit\label{sect:1}}
%=================================================================================================================

\noindent To simplify the discussion, let us consider a one-dimensional problem, and later on 
extend our conclusions to three dimensions. Let us consider a one-dimensional lattice of spacing $a$. 
Each site can be occupied by either a crowder or a tagged particle. We denote with the binary 
variables $m_{i}(k)$ and $n_{i}(k)$ the {\em occupancies} of site $i$ at time $t = k \Delta t$ for the 
tagged and crowding particles, respectively.  Hence $m_{i}(k),n_{i}(k)$ can be either zero or one 
depending on whether site $i$ is occupied or not by the respective particle (see Fig.~\ref{fig:exclud}).\\
\indent The stochastic  process that governs jumps of the tagged particles can be cast in the following form 
\begin{widetext}
\begin{eqnarray}
\label{e:tagged}
m_{i}(k+1)- m_{i}(k) &=& z^{+}_{i-1} m_{i-1}(k)[1-m_{i}(k)] [1-n_{i}(k)] 
                         + z^{-}_{i+1} m_{i+1}(k)[1-m_{i}(k)] [1-n_{i}(k)] \nonumber \\
                      && - z^{+}_{i} m_{i}(k)[1-m_{i+1}(k)][1-n_{i+1}(k)] 
                         - z^{-}_{i} m_{i}(k)[1-m_{i-1}(k)][1-n_{i-1}(k)]
\end{eqnarray}
\end{widetext}
An equivalent equation can be written for the evolution of the crowders' occupancies $n_{j}(\cdot)$. 
Eq.~\eqref{e:tagged}, and its  analogue for species $n_{i}(k)$, can be regarded as the update 
rule for a simple Monte Carlo process. Let us emphasize again that $m_{i}(k)$ and $n_{i}(k)$ are binary variables, 
either zero or one, that specify the occupancy of site $i$. If the target site is occupied by either a crowder 
or a tagged particle, the move cannot occur. 
The quantities $z^{\pm}_{i}$  are variables that take the value $0$ or $1$  
depending on a random number $\xi_i$  uniformly distributed between $0$ and $1$. 
By considering homogeneous jump probabilities, $q^{\pm}_{j}=q$   for $j=i, i \pm 1$, one can formally write
\begin{eqnarray}
\label{e:eta}
z^{+}_{i-1} &=& \theta(\xi_i)-\theta(\xi_i-q) \nonumber \\
z^{-}_{i+1} &=& \theta(\xi_i-q)-\theta(\xi_i-2q) \nonumber \\
z^{+}_{i} &=& \theta(\xi_i-2q)-\theta(\xi_i-3q) \nonumber \\
z^{-}_{i} &=& \theta(\xi_i-3q)-\theta(\xi_i-4q) 
\end{eqnarray}
where $\theta(x)$ is the Heaviside step function and  we are assuming  $q\le 1/4$. 
Eqs.~\eqref{e:eta} entail $\langle z^{\pm}_{j} \rangle = q$, where  $\langle \dots \rangle$  
denotes an average over many values of $\xi_i$ for a fixed configuration $\{ n_{i}, m_{i} \}$. 
The above process is entirely determined by the jump probabilities $q$, which we here assume constant 
and homogeneous. \\
\indent A (discrete-time) master equation can be obtained by averaging over many Monte Carlo realizations
performed according to the rule~\eqref{e:tagged} and starting from the same initial condition
(we denote this average by $\langle\langle\dots\rangle\rangle$). 
Introducing the one-body occupancy probabilities   
\begin{eqnarray}
\rho_{i}(k) = \langle \langle m_{i}(k) \rangle \rangle \label{e:Poneb}\\
\phi_{i}(k) = \langle \langle n_{i}(k) \rangle \rangle
\label{e:Ptwob_tagged}
\end{eqnarray}
and assuming a mean-field factorization for the two-body and three-body correlations, one eventually ends up with
\begin{widetext}
\begin{eqnarray}
\label{e:SEPMEq_1}
 \rho_{i}(k+1) - \rho_{i}(k) &=& q \left(\rho_{i-1}(k) + \rho_{i+1}(k) \right) \left[1 - \rho_{i}(k) \right] \left[1 - \phi_{i}(k) \right]\nonumber \\
&-&q \,\rho_{i}(k) \left[2 - \left(\rho_{i-1}(k) + \rho_{i+1}(k) \right) - \left(\phi_{i-1}(k) + \phi_{i+1}(k) \right) + \phi_{i+1}(k)\rho_{i+1}(k)+\phi_{i-1}(k)\rho_{i-1}(k) \right] \nonumber \\
\phi_{i}(k+1) - \phi_{i}(k) &=& w \left(\phi_{i-1}(k) + \phi_{i+1}(k) \right) \left[1 - \phi_{i}(k) \right] \left[1 - \rho_{i}(k) \right]\nonumber \\
                            &-& w \,\phi_{i}(k)\left[2 - \left(\phi_{i-1}(k) + \phi_{i+1}(k) \right) - \left(\rho_{i-1}(k) + \rho_{i+1}(k) \right) +\phi_{i+1}(k)\rho_{i+1}(k)+\phi_{i-1}(k)\rho_{i-1}(k)\right] 
\end{eqnarray}
\end{widetext}
where $w$ denotes the jump probability associated with crowders' motion. To proceed in the analysis, we assume that 
the concentration of tagged particles is small, $\rho_i\ll 1$. We therefore  approximate eqs.~\eqref{e:SEPMEq_1} as
\begin{eqnarray}
\label{e:SEPMEq_fin}
\rho_{i}(k+1) - \rho_{i}(k) &=& q \left(\rho_{i-1}(k) + \rho_{i+1}(k) \right) \left[1 - \phi_{i}(k) \right]\nonumber \\
&-&q \rho_{i}(k)[2 - \left(\phi_{i-1}(k) + \phi_{i+1}(k) \right)] \nonumber \\
\phi_{i}(k+1) - \phi_{i}(k) &=& w \left( \phi_{i-1}(k) + \phi_{i+1}(k) - 2 \phi_{i}(k)  \right)			     
\end{eqnarray}
Note that the microscopic exclusion constraint is lost in the equation for $\phi_{i}$, the crowders occupancy 
probability. Tagged particles are in fact highly diluted and thus cause a modest 
(negligible, at the considered order of approximation) interference to the     
diffusive motion of the crowders.\\
\indent Let us now move to the continuum. We do so formally by letting
\begin{equation}
\rho(x,t) = \lim_{a,\Delta t\to 0} \rho_{i}(k), \qquad \phi(x,t) = \lim_{a,\Delta t\to 0} \phi_{i}(k).
\end{equation}
In addition we must require 
\begin{eqnarray}
\lim_{a,\Delta t\to0} \frac{q a^{2}}{\Delta t} = D_{\rho} \label{e:diff_tag}\\
\lim_{a,\Delta t\to0} \frac{w a^2}{\Delta t} = D_{\phi} \label{e:diff_crowd}
\end{eqnarray}
where  $D_{\rho}$ and $D_{\phi}$ denote the diffusion coefficients of the tagged particles and the crowders, respectively. 
If the former are a labeled subset of a single population of interacting agents, then $q=w$, which in turn implies
$D_{\rho}=D_{\phi}$.
Making us of the above definitions, one readily obtains the continuum limit of eqs.~\eqref{e:SEPMEq_fin} 
\begin{eqnarray}
\label{sistema}
\frac{\partial\rho}{\partial t}&=& \frac{\partial^2}{\partial x^2} \left[ D_{\rho}(1-\phi)\rho\right] +
                            2 D_{\rho} \frac{\partial}{\partial x} \left( \rho\frac{\partial\phi}{\partial x} \right) \nonumber \\
\frac{\partial \phi}{\partial t}&=& D_{\phi} \frac{\partial^2\phi}{\partial x^2} 
\end{eqnarray}
We note that the equations (\ref{sistema}) also govern the evolution of the particles concentrations provided the probabilities 
$\rho$ and $\phi$  are replaced by their maximum values, {\em i.e.} the inverse specific volumes of the particles. 
In the following, the densities $\phi$ and $\rho$ are expressed in units of their corresponding maximum values.

The mean-field density of crowders $\phi$ evolves in time following a standard diffusion equation.
The density $\rho$ obeys instead a nonlinear equation with drift, which bears the signature of the point-like
excluded volume rules imposed at the microscopic level. As we shall demonstrate in the following, the drift term is eventually 
responsible for the  emergence of transient sub-diffusive and super-diffusive dynamics, reflecting the specificity of the initial 
condition selected. The above derivation applies to one spatial dimension, but the result can be readily extended to higher dimensions. 
In the appendix we give an alternative derivation of eqs.\eqref{sistema}, following a perturbative calculation inspired by van Kampen system size expansion~\cite{VanKampen:2011vs}. 
Notice that the equation for the evolution of $\rho$  has also been derived in Ref.~\cite{Seki:2012ux} 
for a constant non-homogeneous background field $\phi(x)$.

%===================================================================================================================================================
%  SECTION 2
%
\section{Sub- and Super-diffusive transients\label{sect:2}}
%===================================================================================================================================================

\begin{figure}[t!]
\centering
\subfigure{\label{fig:label1}\includegraphics[width=\columnwidth]{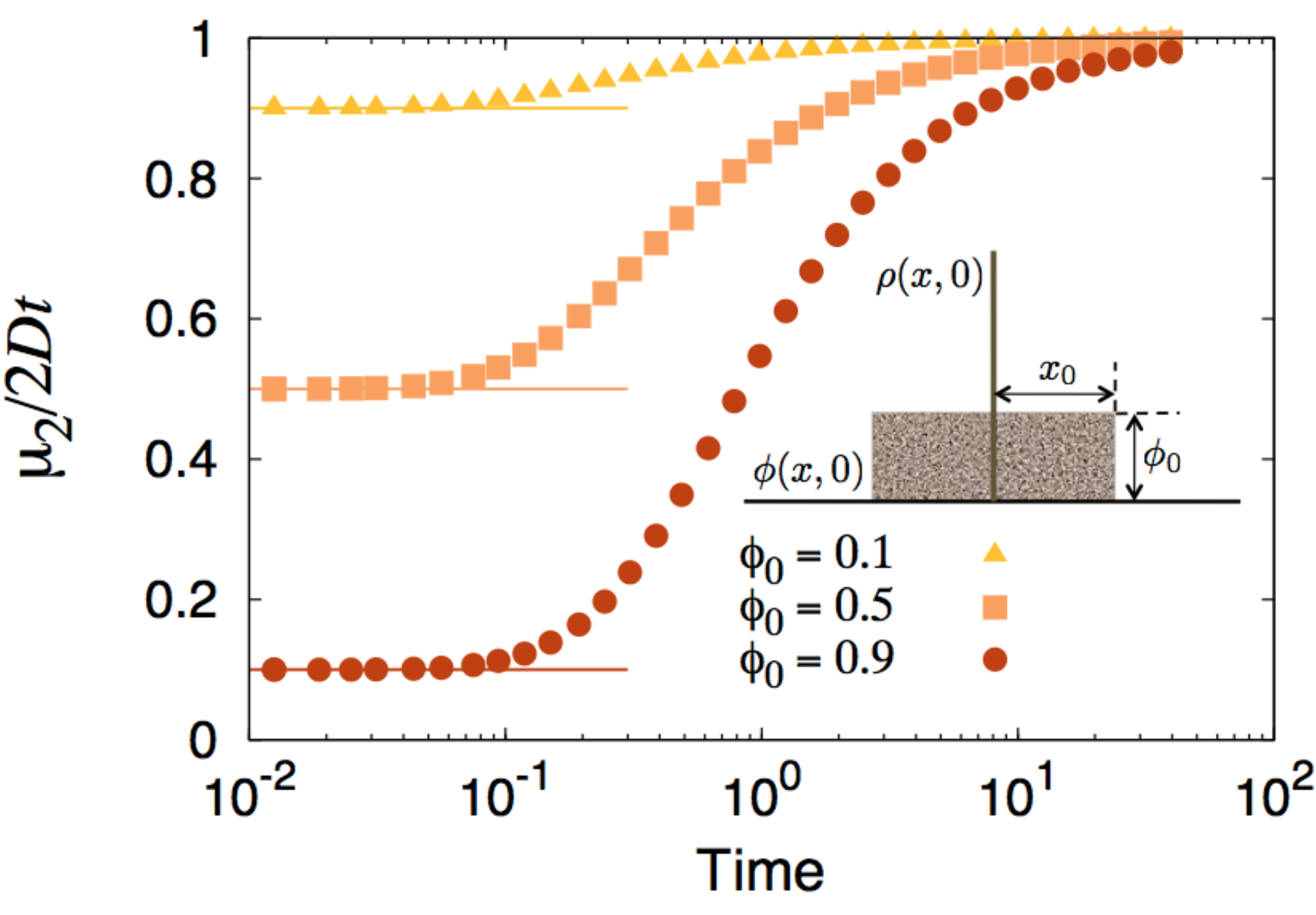}}
\subfigure{\label{fig:label2}\includegraphics[width=\columnwidth]{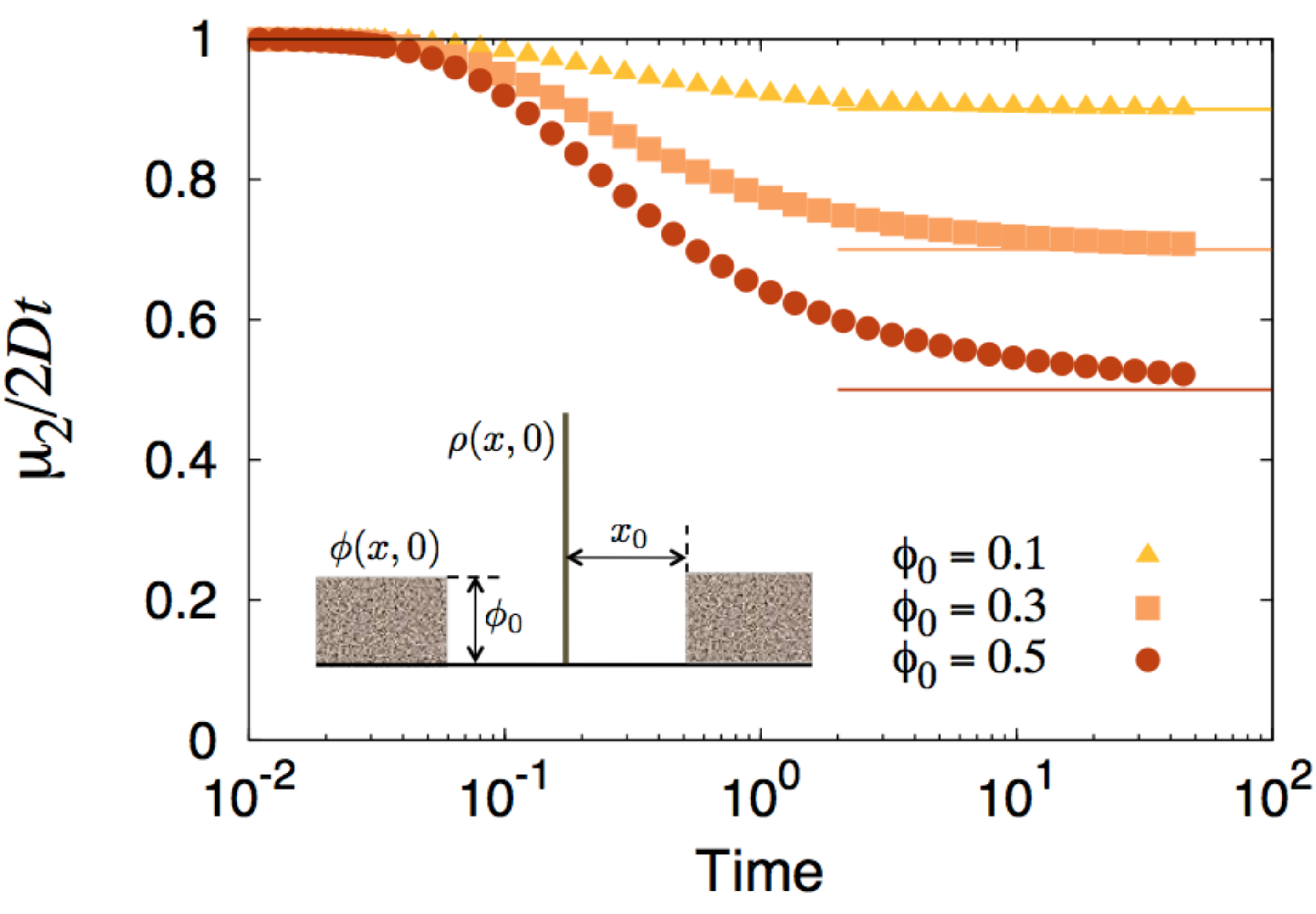}}
\caption{Rescaled mean square displacement $\mu_2/2Dt$ as a function of time, expressed  
in units of $\tau=x_0^2/2D$, for different crowding strengths $\phi_{0}$. 
Upper panel: initial condition of the type~\eqref{condizioniiniziali}, originating a  super-diffusive transient.  
Lower panel  initial condition~\eqref{condizioniiniziali1}, causing a sub-diffusive  transient.  
The parameter $\tau$ quantifies the lifetime of the observed dynamical transients. }
\label{msd}
\end{figure}

\noindent In order to monitor the time-evolution of the tagged species, 
we introduce the mean square displacement (MSD) $\mu_2(t)$
\begin{equation}
\label{e:mu2def}
\mu_2(t) = \int x^2 \rho (x,t) dx - \left(\int x \rho (x,t) dx \right)^2
\end{equation}
It is well known that the mean square displacement 
scales linearly with time for unobstructed diffusion. 
As detailed in the introduction, a sub-linear growth of the MSD is often interpreted 
as a direct manifestation of the microscopic competition for available space in crowded media. 
As we shall prove in the following, this is an overly simplistic picture, as   
more complex scenarios can easily be obtained by direct integration of eqs.~\eqref{sistema}, where
nonlinear MSD emerge only as {\em  transient} regimes.     
We are in particular interested in a specific class of initial condition, symmetric in the domain of definition,
so  that $\int x \rho (x,t) dx=0$, and the MSD equals the second moment of 
the tagged particles distribution $\rho$.\\
\indent At time $t=0$, the tagged species is supposed to be localized at the origin. 
In formulae, $\rho(x,0)=\delta (x)$, where $\rho(\cdot)$ is Dirac delta. Let us first  
assume that the crowders initially populate a compact domain, centered at the origin. 
The initial distribution is of the water-bag type, that is
\begin{equation}
\label{condizioniiniziali}
\phi(x,0)= \phi_0 \left[ \theta(x+x_0) - \theta(x-x_0) \right] 
\end{equation}
where $x_0$ is the semi-width of the water-bag, $\phi_0 \in [0, 1]$ and $\theta(\cdot)$ is the Heaviside function
(see inset in the upper panel of Fig.~\ref{msd}). From here on, as a further simplification, 
we assume $D_{\rho}=D_{\phi}=D$. \\
Fig.~\ref{msd} shows  the rescaled MSD as a function of time, as obtained by numerically 
integrating eqs.~\eqref{sistema} through an explicit Euler discretization scheme. 
At short times, the tagged species is immersed in the almost uniform sea of surrounding crowders. 
Since $\phi$ is approximately constant,  the tagged particles diffuse normally. 
In fact, the first equation of~\eqref{sistema} simplifies for $\phi(x,t)=\phi_0$, yielding a 
standard diffusion equation for $\rho(x,t)$ with an effective diffusion coefficient equal to $D(1-\phi_0)$. \\
\indent Let us now focus  on the long-time dynamics. The crowders are spread over the one-dimensional
support, which we imagine open but very large so as to neglect boundary effects. 
The  density $\phi$ is consequently small and its contribution can be
neglected in the Fokker-Planck equation for the evolution of $\rho$. Again, we recover normal diffusion
with diffusion coefficient $D$. 
In short, the rescaled MSD $\mu_2/2Dt$ is close to  $(1-\phi_0)$, at short times, and converges 
asymptotically to $1$.  The two regimes of normal diffusion appear bridged by a super-diffusive crossover. \\
\indent As anticipated, a monotonic continuous curve is found to smoothly link the two trivial limiting solutions at 
short and long times. It is remarkable, and to some extent counter-intuitive, that a super-diffusive transient is found 
in a model accounting for crowding. The latter self-consistently accommodates for the microscopic excluded volume interactions 
among diffusing agents, a process that is customarily believed to display sub-diffusive spread of concentrations. 
We observe that the time duration of the  super-diffusive transient increases quadratically with $x_0$, the width of the initial 
water bag.\\
\indent To understand the origin of the observed dynamics, let us go back to the Fokker-Planck equation 
and focus on the drift term, namely $-\partial  \left( \rho(x,t) v(x,t) \right )/\partial x$,
where $v(x,t)=-\partial \phi(x,t) / \partial x $. This is an effective velocity field, induced by the crowders, 
that acts as a systematic bias in the evolution of the density $\rho$. Initially, $\rho$ is subject to a zero 
velocity field, as $\phi(x,t) \simeq \phi_0$, for all values of $x$ where $\rho$ is non-zero. 
Then, after a time of the order of $\tau \propto x_{0}^{2}$, the support of $\rho$ extends to a domain where 
it is no longer possible to assume $\phi(x,t)$ constant. In particular, $\partial \phi(x,t)/\partial x<0$, for $x>0$, 
which implies $v(x,t)>0$. 
A similar reasoning allows us to conclude that $v(x,t)<0$, when $x<0$.  The mean field drift, which ultimately stems 
from the microscopic competition for space between crowders and tagged particles, pulls the distribution $\rho$ away from the origin, 
stretching the right (left) tail towards the direction of positive (negative) $x$. This leads to the (apparent) super-diffusive 
transient shown in Fig.~\ref{msd}.\\
\indent A dual situation can be imagined  yielding a sub-diffusive transient. 
To this end, let us consider a one-dimensional domain of size $2 L$ and assume that $\phi$ is 
therein uniformly distributed. At time $t=0$, the crowders that populate a segment of width $2 x_0$, 
centered around the origin, are removed from the system. This amounts to considering an initial 
distribution for $\phi$ that is a superposition of two water-bags (see inset in the bottom panel of Fig.~\ref{msd}). 
In formulae, 
\begin{equation}
\label{condizioniiniziali1}
\phi(x,0) = \phi_0 \left[ 1- \theta(x+x_0)  + \theta(x-x_0) \right] 
\end{equation}
At short times, the diffusion of tagged particles inserted at the origin is not affected by the crowders. 
The rescaled MSD $\mu_2/2Dt$ is hence approximately equal to one and stays constant over a finite  
time window of order $x_0^2/D$. Eventually, the crowders have approximately relaxed to the uniform distribution characterized by the asymptotic 
concentration $\phi_L= \phi_0 (1-x_0/L)$. At this stage, the tagged particles find themselves 
in a uniformly crowded medium with a reduced diffusion coefficient $D(1- \phi_L)$. The bottom panel of Fig.~\ref{msd} confirms 
our reasoning, as the rescaled MSD $\mu_2/2Dt$ is seen to decrease monotonously, 
interpolating between the initial plateau $\mu_{2} = 1$  
and the final value $\mu_{2} = (1- \phi_L)<1$. In this case, one thus observes a sub-diffusive crossover. 
Here, the background density acts as an external potential that contrasts the spreading of the distribution. 
In this case, in fact,  
$\partial \phi(x,t) / \partial x>0$, for $x>0$, which implies $v(x,t)<0$, {\em i.e.} 
a drift that opposes the diffusive thrust to delocalization. 
%
%%%%%%%%%%%%%%%%%%%%%%%%%%%%%%%%%%%%%%%%%%%%%%%%%%%%%%%%%%%%%%%%%%%%%%%%%%%%%%%%%%%%%%%%%%%%%%%%%%%%%%%%%%%%%%%%%%%%
%%%%%%%%%%%%%%%%%%%%%%%%%%%%%%%%%%%%%%%%%%%%%%%%%%%%%%%%%%%%%%%%%%%%%%%%%%%%%%%%%%%%%%%%%%%%%%%%%%%%%%%%%%%%%%%%%%%%

%================================================================================================================================================
% CONCLUSIONS
%
\section{Conclusion\label{sect:3}}
%================================================================================================================================================

\noindent The study of molecular diffusion under crowded conditions is an interesting topic of investigation, 
particularly crucial for its applications to cellular biology. At high density, particles diffusion is impeded 
and excluded-volume effects become complex and cannot be ignored. 
In this paper we have considered the diffusive dynamics of an ensemble of inert particles, 
the {\em tagged} species, immersed in a crowded background of co-evolving agents. 
This is a quite general scenario, which can be invoked to  describe different experimental conditions. 
The tagged particles are assumed to be sufficiently diluted, a working hypothesis that allows us to neglect 
their feedback on the crowders. As a consequence, the continuum density of the tagged species is 
governed by a nonlinear Fokker-Planck equation with non-homogeneous drift and diffusion coefficients,
which are self-consistently determined by the time-dependent concentration of crowders. 
In the background, the crowders are undisturbed and undergo normal diffusion. \\
\indent Working within this framework, we have shown that transient sub-diffusive and super-diffusive regimes 
can emerge, depending  on the specific initial condition. When the crowders are uniformly dispersed in the 
container, but removed from an isolated patch where the tagged species is initially confined, a sub-diffusive 
scaling for the mean square displacement is observed. This crossover regime persists within a finite, possibly very long
time window.
We observe that crowding is rather often associated with anomalous slowing down of transport, {\em i.e.}
sub-diffusion. It is therefore surprising that the dynamical 
interference between crowders and tagged particles can result in super-diffusive dynamics 
for certain choices of the initial condition. Imagine that the tagged agents are trapped inside a 
uniform patch of crowders inside a much larger, otherwise empty container. 
In this case, point-like excluded volume interactions of the kind considered here results in a drift term in the 
Fokker-Planck equation for the tagged particles, which accelerates their spread as compared to diffusion. 
This condition can be easily recreated in laboratory experiments, by initially confining the particles, 
including those whose evolution is to be tracked, within a finite portion of the available space. \\
\indent The presence of super-diffusion in a toy model of percolation with mobile obstacles has been 
previously observed by Stauffer and collaborators~\cite{Stauffer:2008bx}. We get to the same  conclusion in our paper by
studying a nonlinear mean-field model, derived from first principles, which generalizes standard diffusion to transport in 
the presence of mobile crowders.  Our findings suggest that both super-diffusion and sub-diffusion transients can occur
beyond the idealized diluted limit, strongly warning 
against the simplistic identification of crowding with anomalous transport, in particular sub-diffusion. 
As a final comment, we also stress that, for the sake of simplicity, the analysis is here carried out in one 
spatial dimension. However, our conclusions are general and 
can be readily extended to higher spatial dimensions.

%%%%%%%%%%%%%%%%%%%%%%%%%%%%%%%%%%%%%%%%%%%%%  REFERENCES %%%%%%%%%%%%%%%%%%%%%%%%%%%%%%%%%%%%%%%%%%%%%%%%%%%%%%%%%%%%%%%%%%%%%%%%%%%%%%%%%%
%\bibliography{crowding}
%

%
%%%%%%%%%%%%%%%%%%%%%%%%%%%%%%%%%%%%%%%%%%%%%%%%%%%%%%%%%%%%%%%%%%%%%%%%%%%%%%%%%%%%%%%%%%%%%%%%%%%%%%%%%%%%%%%%%%%%%%%%%%%%%%%%%%%%%%%%%%%%

\appendix
\section{Alternative derivation of the model using a coarse grained picture\label{sect:app}} 

\noindent In this Appendix we discuss an alternative derivation of model \eqref{sistema}, which assumes a 
coarse-grained decription of the scrutinized problem. The derivation follows a different philosphy: it 
is here carried out in one dimension, but readily generalizes to the relevant $d=3$ setting. We consider the physical space to be partitioned in $\Omega$ patches, also called {\em urns}.
Each patch has a maximum carrying capacity -  it can be filled with $N$ particles at most. Labelling $m_i$ the 
number of tagged particles contained in urn $i$, and with $n_i$ the corresponding number of crowders, one can write:
\begin{equation*}
n_i+m_i+v_i=N  \ \ \forall i
\end{equation*} 
where $v_i$ stands for the number of vacancies, the empty cases in patch $i$ that can be eventually filled by incoming particles. 
The excluded-volume prescription is here implemented by  requiring that particles can move only into the nearest-neighbor 
patches that exhibit vacancies, as exemplified by the following chemical reactions
\begin{equation}
\label{chemical}
\begin{split}
\mathcal{M}_i+V_j\stackrel{\frac{\delta}{z\Omega}}{\longrightarrow} \mathcal{M}_j+V_i
\\\mathcal{N}_i+V_j\stackrel{\frac{\delta}{z\Omega}}{\longrightarrow} \mathcal{N}_j+V_i
\end{split}
\end{equation}
Here $z$ is the number of nearest-neighbor patches and $\mathcal{M}_i,\mathcal{N}_i,V_i$ 
are respectively a particle of type $\mathcal{M}$ (the tagged particles), 
of type $\mathcal{N}$ (the crowders) or a vacancy belonging to the $i$-patch.\\
\indent This is stochastic process governed, under the Markov hypothesis, by a Master equation for the probability   
$P(\bold{n},\bold{m},t)$ of finding the system in a  given state specified by the $2 \Omega$ dimensional 
vector $(\bold{n},\bold{m})=(n_1,...,n_{\Omega},m_1...,m_{\Omega})$ at time $t$.  The Master equation reads:
\begin{eqnarray}
\label{generale}
&\frac{\partial P(\bold{n},\bold{m},t)}{\partial t}=
\sum_{n \neq n'} [ T(\bold{n},\bold{m}|\bold{n}',\bold{m})P(\bold{n}',\bold{m})+\\ \nonumber
&T(\bold{n},\bold{m}|\bold{n},\bold{m}') P(\bold{n},\bold{m}')
- T(\bold{n},\bold{m}'|\bold{n},\bold{m})P(\bold{n},\bold{m})-\\ \nonumber
&T(\bold{n}',\bold{m}|\bold{n},\bold{m}) P(\bold{n},\bold{m}) ]
\end{eqnarray}
where $T(\bold{a}|\bold{b})$ is the rate of transition from a state $\bold{a}$ to a compatible configuration  $\bold{b}$.
The allowed transitions are those that take place between neighboring patches as dictated by the chemical reactions $(\ref{chemical})$. 
For example, the transition probability associated with the second of equations $(\ref{chemical})$ reads
\begin{equation}
\label{rates}
T(n_i -1,n_j +1|n_i,n_j)=\frac{\delta}{z\Omega}\frac{n_i}{N}\frac{v_j}{N}=
\frac{\delta}{z\Omega}\frac{n_i}{N}(1-\frac{n_j}{N}-\frac{m_j}{N}).
\end{equation}
The transition rates bring into the equation an explicit dependence on the amount of molecules per patch $N$, 
the so-called system size. To proceed in the analysis, we make use of van Kampen system size expansion~\cite{VanKampen:2011vs},
which enables one to separate the site-dependent mean concentration $\phi_i(t)$ from the
corresponding fluctuations $\xi_i$  in the expression of the discrete number density of species $\mathcal{N}$. 
The fluctuations become less influent as the number of the 
agents is increased, an observation which translates in the following van Kampen ansatz:
\begin{equation}
\label{vankampen}
\frac{n_i}{N}(t)=\phi_i(t)+\frac{\xi_i}{\sqrt{N}}.
\end{equation}
In the following we will also assume just one tagged particle, the analysis extending straightforwardly 
to the case where a bunch of diluted particles is assumed to be 
dispersed in the background of crowders.  Since the tagged particle belongs to one of the patches, 
it is convenient to look at the evolution of
\begin{equation*}
 P_k(\bold{n},t)=P(\bold{n},\underbrace{0,0,...,0}_{k-1},1,\underbrace{0,....,0,0}_{\Omega-k},t)
 \end{equation*} 
in the master equation ($\ref{generale}$). $ P_k(\bold{n},t)$ is the probability that the 
target particle be in the $k$-patch, for a particular configuration $\bold{n}$ of species $\mathcal{N}$.
The Master equation can be hence written in the following compact form
\begin{widetext}
\begin{equation}
\label{maestra2}
\begin{split}
\frac{\partial{P_k(\bold{n},t)}}{\partial t}&=
\sum_{i=1}^{\Omega}\sum_{j\in{i-1,i+1}} (\epsilon_j^{-}\epsilon_i^{+}-1)T(n_i -1,n_j +1|n_i,n_j) P_k(n_i,n_j,t)
\\&+\sum_{i=1}^{\Omega} \left\{
-\sum_{j\in{i-1,i+1}}
\frac{\delta}{z\Omega}\frac{1}{N}\bigg(1-\frac{n_j}{N}\bigg)P_k+\sum_{j\in{i-1,i+1}} \frac{\delta}{z\Omega}\frac{1}{N}\bigg(1-\frac{n_k}{N}\bigg)P_j\right\}
\end{split}
\end{equation}
\end{widetext}
where use has been made of the shift operators:
$$\epsilon_i^{\pm}f(....,n_i,.....)=f(....,n_i\pm 1,.....).$$
Under the van Kampen prescription~\cite{VanKampen:2011vs}, one can expand 
the transition rates in power of $1/\sqrt{N}$. For example, equation (\ref{rates}) takes the form
\begin{widetext}
\begin{equation*}
 \begin{split}
T(n_i-1,n_j+1|n_i,n_j)&=\frac{\delta}{z\Omega}\left\{(\phi_i(1-\phi_j))+\frac{1}{\sqrt{N}}\big[\xi_i(1-\phi_j)-\xi_j\phi_i]+\frac{1}{N}[-\xi_i\xi_j-m_j\phi_i] +\frac{1}{N^\frac{3}{2}}\big[-m_j\xi_i\big]\right\}
\end{split}
\end{equation*}
\end{widetext}
and also express the shift operators in terms of differential operators:
\begin{equation*}
(\epsilon_j^{-}\epsilon_i^{+}-1)=\frac{1}{\sqrt{N}}\bigg(\frac{\partial}{\partial \xi_i}-\frac{\partial}{\partial \xi_j}\bigg)+\frac{1}{2N}
\bigg(\frac{\partial}{\partial \xi_i}-\frac{\partial}{\partial \xi_j}\bigg)^2+O(\frac{1}{N^\frac{3}{2}})
\end{equation*}
%
%
%leading to
\begin{widetext}
\begin{eqnarray*}
%\begin{split}
&(\epsilon_j^{-}\epsilon_i^{+}-1)T(n_i -1,n_j +1|n_i,n_j) P(n_i,n_j,t)=\frac{1}{\sqrt{N}}\bigg[\bigg(\frac{\partial}{\partial \xi_i}-\frac{\partial}{\partial \xi_j}\bigg)\bigg(\frac{\delta}{ z\Omega}(\phi_i(1-\phi_j))\Pi_k(\xi,t)\bigg)\bigg]
\\&+
\frac{1}{N}\bigg[\bigg(\frac{\partial}{\partial \xi_i}-\frac{\partial}{\partial \xi_j}\bigg) \bigg(\frac{\delta}{ z\Omega}(\xi_i(1-\phi_j)+\xi_j\phi_i)\Pi_k(\xi,t)\bigg) +\frac{1}{2}\bigg(\frac{\partial}{\partial \xi_i}-\frac{\partial}{\partial \xi_j}\bigg)^2\bigg(\frac{\delta}{ z\Omega}(\phi_i(1-\phi_j))\Pi_k(\xi,t)\bigg)\bigg]+O\bigg(\frac{1}{N^\frac{3}{2}}\bigg).
%%\end{split}
\end{eqnarray*}
\end{widetext}
Notice that $m_i/N$ cannot be approximated as a continuum-like density, the continuum limit being not appropriate 
for the case of a single tracer.\\
\indent We then define a new probability distribution $\Pi_k(\boldsymbol{\xi},\tau)$, 
function of the vector $\boldsymbol{\xi}$ and the 
scaled time $\tau=\frac{t}{N\Omega}$. In terms of the new probability distribution $\Pi_k(\boldsymbol{\xi},\tau)$ the left hand side of (\ref{maestra2}) becomes 
\begin{equation*}
\frac{\partial P_k}{\partial{t}}=-\frac{1}{\sqrt{N}\Omega}\sum_{i=1}^{\Omega}\frac{\partial\Pi_k}{\partial\xi_i}\dot{\phi}_i+ \frac{1}{N\Omega}\frac{\partial\Pi_k}{\partial t}.
\end{equation*}
The leading order contribution in ($\frac{1}{\sqrt{N}}$) gives:
\begin{equation}
-\frac{1}{\Omega}\sum_{i=1}^{\Omega}\frac{\partial\Pi_k}{\partial\xi_i}\dot{\phi_i}=\frac{\delta}{z\Omega}\sum_{i=1}^{\Omega}\sum_{j\in\{i-1,i+1\}}\phi_i(1-\phi_j)(\frac{\partial\Pi_k}{\partial\xi_i}-\frac{\partial\Pi_k}{\partial\xi_j})
\end{equation}
which yields
\begin{equation}
\sum_{i=1}^{\Omega}\frac{\partial\Pi_k}{\partial\xi_i}\dot{\phi_i}=\frac{\delta}{z}\sum_{i=1}^{\Omega}-\frac{\partial\Pi_k}{\partial\xi_i}(2\phi_i-\phi_{i-1}-\phi_{i+1})
\end{equation}
and finally:
\begin{equation}
\dot{\phi}_i=\frac{\delta}{2}\Delta\phi_i
\end{equation}
where $\Delta$ is the discrete Laplacian operator defined as 
$\Delta\phi_i=\frac{2}{z}\sum_{j\in i}(\phi_j-\phi_i)$, where $\sum_{j\in i}$ means a summation over the sites, $j$, which are nearest-neighbors of site $i$. By taking the size of the patches to zero, one 
recovers the standard diffusion equation for species $\phi$, in agreement with the result reported in 
the main body of the paper. Consider now the following identities:
\begin{widetext} 
 \begin{equation*}
\begin{split}
&\frac{\delta}{z}\sum_{i=1}^{\Omega}\sum_{j\in\{i-1,i+1\}}(\frac{\partial}{\partial\xi_i}-\frac{\partial}{\partial\xi_j})(\xi_i(1-\phi_j)-\xi_j\phi_i)\Pi_k
\\
&=\frac{\delta}{z}\sum_{i=1}^{\Omega}(\frac{\partial}{\partial\xi_i}-\frac{\partial}{\partial\xi_{i-1}})(\xi_i(1-\phi_{i-1})-\xi_{i-1}\phi_i)\Pi_k+  \frac{\delta}{z}\sum_{i=1}^{\Omega}(\frac{\partial}{\partial\xi_i}-\frac{\partial}{\partial\xi_{i+1}})(\xi_i(1-\phi_{i+1})-\xi_{i+1}\phi_i)\Pi_k
\\
&=\frac{\delta}{z}\sum_{i=1}^{\Omega}\frac{\partial}{\partial \xi_i}
\bigg((\xi_i(1-\phi_{i-1})-\xi_{i-1}\phi_i)\Pi_k +
(\xi_i(1-\phi_{i+1})-\xi_{i+1}\phi_i)\Pi_k\bigg)
\\
&-\frac{\delta}{z}\sum_{i=1}^{\Omega}\frac{\partial}{\partial \xi_i}(\xi_{i+1}(1-\phi_i)-\xi_i\phi_{i+1})\Pi_k
-\frac{\delta}{z}\sum_{i=1}^{\Omega}\frac{\partial}{\partial \xi_i}(\xi_{i-1}(1-\phi_i)-\xi_i\phi_{i-1})\Pi_k 
\\&=\frac{\delta}{2}\sum_{i=1}^{\Omega}-\Delta\xi_i\Pi_k
\end{split}
\end{equation*}
\end{widetext}
and
\begin{widetext}
\begin{equation*}
\begin{split}
&\frac{\delta}{z}\sum_{i=1}^{\Omega}\sum_{j\in\{i-1,i+1\}}\bigg(\frac{\partial}{\partial\xi_i}-\frac{\partial}{\partial\xi_j}\bigg)^2
(\phi_i(1-\phi_j))\Pi_k
\\
&=\frac{\delta}{z}\sum_{i=1}^{\Omega}\frac{\partial^2}{\partial \xi_i^2}\bigg(\phi_i(1-\phi_{i-1})\Pi_k\bigg)+
\frac{\partial^2}{\partial \xi_i^2}\bigg(\phi_i(1-\phi_{i+1})\Pi_k\bigg)
\frac{\partial^2}{\partial \xi_{i+1}^2}
\bigg(\phi_i(1-\phi_{i+1})\Pi_k\bigg)+\frac{\partial^2}{\partial \xi_{i-1}^2}
\bigg(\phi_i(1-\phi_{i-1})\Pi_k\bigg)
\\
&-\frac{2\partial^2}{\partial\xi_i\partial\xi_{i-1}}\bigg(\phi_i(1-\phi_{i-1})\Pi_k\bigg)-\frac{2\partial^2}
{\partial\xi_i\partial\xi_{i+1}}\bigg(\phi_i(1-\phi_{i+1})\Pi_k\bigg)
\\
&=\frac{\delta}{z}\sum_{i=1}^{\Omega}\frac{\partial^2}{\delta\xi_i^2}(2\phi_i+\phi_{i-1}+\phi_{i+1}-2\phi_i(\phi_{i+1}+\phi_{i-1}))\Pi_k
+\frac{\partial^2}{\partial\xi_i\partial\xi_{i-1}}\bigg(-2\phi_i(1-\phi_{i-1})\bigg)\Pi_k\\
&+\frac{\partial^2}{\partial\xi_i\partial\xi_{i+1}}\bigg(-2\phi_i(1-\phi_{i+1})\bigg)\Pi_k
\end{split}
\end{equation*}
\end{widetext} 
Making use of the above relations, at the next to next-to-leading corrections one eventually gets
\begin{widetext}
\begin{eqnarray}
\begin{split}
\frac{\partial\Pi_k}{\partial t}&=\frac{\delta}{2}\sum_{i=1}^{\Omega}\frac{\partial}{\partial\xi_i}\bigg(-\Delta\xi_i\Pi_k\bigg)
+\frac{\delta}{2z}\sum_{i=1}^{\Omega}\sum_{i=i-1}^{i+1}\frac{\partial}{\partial \xi_i}\frac{\partial}{\partial
\xi_j}\bigg(B_{i,j}\Pi_k\bigg)\\
&+\frac{\delta}{z}\bigg((1-\phi_k)\Pi_{k-1}-(2-\phi_{k+1}-\phi_{k-1})\Pi_k+(1-\phi_k)\Pi_{k+1}\bigg)
\nonumber
\end{split}
\label{secondoordine}
\end{eqnarray}
\end{widetext}
Here $B$  represents  the diffusion matrix, whose entries are
\begin{equation}
\begin {split} B&_{i,i}=2\phi_i+\phi_{i-1}+\phi_{i+1}-2\phi_i(\phi_{i+1}+\phi_{i-1})
\\B&_{i,i-1}=(-2\phi_i(1-\phi_{i-1}))
\\B&_{i,i+1}=(-2\phi_i(1-\phi_{i+1})).
\end{split}
\end{equation}.
To provide a mean-field description of the inspected problem, 
we consider the probability function of the tagged agent integrated over the fluctuations 
of the $\mathcal{N}$-particles. In formulae: 
$$\rho_k(t)=\int \Pi_k d\boldsymbol{\xi}$$ 
whose evolution is governed by
\begin{eqnarray*}
\frac{\partial\rho_k}{\partial t}&=&\frac{\delta}{z}\big((1-\phi_k)\rho_{k-1}-(2-\phi_{k+1}-\phi_{k-1})\rho_k \\ \nonumber
&+&(1-\phi_k)\rho_{k+1}\big) =\frac{\delta}{2}(\Delta\rho_k-\phi_k\Delta\rho_k+\rho_k\Delta\phi_k)
\end{eqnarray*}
The the last expression involves the discrete laplacian $\Delta$ defined above.  In the continuum limit, 
and considering a straightforward generalization to higher dimensions, one gets 
\begin{equation*}
\frac{\partial\rho(x,t)}{\partial t}=D_{\rho}(1-\phi(x,t))\nabla^2\rho(x,t)+D_{\rho} \rho(x,t)\nabla^2\phi(x,t)
\end{equation*}
where $D_{\rho}$ is the diffusion coefficient of the tagged particle. 
One can finally write the non-linear equation for 
$\rho$ as a Fokker Plank equation: 
\begin{eqnarray*}
\label{fok}
\frac{\partial\rho(x,t)}{\partial t}&=\nabla^2\bigg(D(1-\phi(x,t))\rho(x,t)\bigg)\\
&+2D\nabla\bigg(\rho(x,t)\nabla\phi(x,t)\bigg) \nonumber
\end{eqnarray*}
Hence, by neglecting the role of fluctuations, which amounts to operating in the mean-field limit, 
a nonlinear partial differential equation is found for the density of the tagged species, coupled to a standard diffusion equation for the background density:
\begin{widetext}
\begin{equation}
\label{eq_fin}
\begin{cases}
 \frac{\partial \phi(x,t)}{\partial t}=D\nabla^2\phi(x,t) \\
\frac{\partial\rho(x,t)}{\partial t}=\nabla^2\bigg(D(1-\phi(x,t))\rho(x,t)\bigg)+2D\nabla\bigg(\rho(x,t)\nabla\phi(x,t)\bigg)
\end{cases}
\end{equation}
\end{widetext}
This system constitutes the generalization of model~\eqref{sistema}) to higher dimensions. It is worth emphasising that the second of eqs. (\ref{eq_fin}) can be also cast in the alternative form:

\begin{equation*}
\frac{\partial\rho(x,t)}{\partial t}=\nabla\cdot \left(D\left(1-\phi(x,t)\right)\nabla\rho(x,t)+D\rho(x,t)\nabla\phi(x,t)\right)
\end{equation*}

\end{document}